\documentclass[aps,prd,nofootinbib,twocolumn,superscriptaddress]{revtex4}

\usepackage{graphicx}
\usepackage{amsmath}

\newcommand{\ie}{{\it i.e., }}

\newcommand{\be}{\begin{equation}}
\newcommand{\ee}{\end{equation}}
\newcommand{\br}{\begin{eqnarray}}
\newcommand{\bea}{\begin{eqnarray}}
\newcommand{\eea}{\end{eqnarray}}
\newcommand{\er}{\end{eqnarray}}
\newcommand{\ba}{\begin{array}}
\newcommand{\ea}{\end{array}}
\newcommand{\bi}{\begin{itemize}}
\newcommand{\ei}{\end{itemize}}
\newcommand{\bn}{\begin{enumerate}}
\newcommand{\en}{\end{enumerate}}
\newcommand{\bc}{\begin{center}}
\newcommand{\ec}{\end{center}}

\newcommand{\beq}{\begin{equation}}
\newcommand{\eeq}{\end{equation}}

\newcommand{\sd}{\sigma_{\!\scriptscriptstyle{S}}}
\newcommand{\UB}{{\scriptscriptstyle{U\!B}}}
\newcommand{\mo}{{\scriptscriptstyle{-1}}}
\newcommand{\HWW}{HW^{\!\scriptscriptstyle{+}}\!W^{\!\scriptscriptstyle{-}}\!\!}
 
\newcommand{\td}{\mathrm{d}}

\newcommand{\slD}{\kern2pt/\kern-8pt D}
\newcommand{\slP}{P\kern-7pt/\kern2pt}
\newcommand{\sla}{a\kern-5pt\raise1pt\hbox{$\scriptstyle/$}\kern1pt}
\newcommand{\sle}{/\kern-5pt\epsilon}
\newcommand{\slk}{/\kern-6pt k}
\newcommand{\sll}{/\kern-4pt l}
\newcommand{\slp}{p\kern-5pt/}
\newcommand{\slq}{q\kern-5.0pt/}
\newcommand{\sls}{s\kern-4.5pt/}
\newcommand{\slv}{v\kern-5pt\raise1pt\hbox{$\scriptstyle/$}\kern1pt}
\newcommand{\sleps}{\epsilon\kern-5.5pt/}
\newcommand{\slpartial}{\partial\kern-5pt/}

\newcommand{\bspace}{\!\!\!\!}
\def\met{\mbox{$E{\bspace}/_{T}$}}

\newcommand{\gsim}{\lower.7ex\hbox{$\;\stackrel{\textstyle>}{\sim}\;$}}
\newcommand{\lsim}{\lower.7ex\hbox{$\;\stackrel{\textstyle<}{\sim}\;$}}

\def\mysection#1{{\bf #1.} }

\begin{document}

\title{Anomalous Higgs-boson coupling effects in $\HWW$   production 
 at the LHC.}

\author{Emidio Gabrielli\footnote{
On leave of absence from Department of Physics, 
University of Trieste, Strada Costiera 11, I-34151 Trieste, Italy.}}
\affiliation{National Institute of Chemical Physics and Biophysics, R\"avala 10, 10143 Tallinn, Estonia}
\affiliation{INFN, Sezione di Trieste, Via Valerio 2, 34127 Trieste, Italy}
 
\author{Matti Heikinheimo}
\affiliation{National Institute of Chemical Physics and Biophysics, R\"avala 10, 10143 Tallinn, Estonia}

\author{Luca Marzola}
\affiliation{Laboratory of Theoretical Physics, Institute of Physics, University of Tartu, T\"ahe 4, 51010 Tartu, Estonia}

\author{Barbara Mele}
\affiliation{INFN, Sezione di Roma, Piazzale Aldo Moro 2, 00185 Roma, Italy}

\author{Christian Spethmann}
\affiliation{National Institute of Chemical Physics and Biophysics, R\"avala 10, 10143 Tallinn, Estonia}

\author{Hardi Veerm\"ae}
\affiliation{National Institute of Chemical Physics and Biophysics, R\"avala 10, 10143 Tallinn, Estonia}
\affiliation{Laboratory of Theoretical Physics, Institute of Physics, University of Tartu, T\"ahe 4, 51010 Tartu, Estonia}

\date{\today}

\begin{abstract}
We study the LHC associated production of a Higgs boson and a  $W^{\!\scriptscriptstyle{+}}\!W^{\!\scriptscriptstyle{-}}\!$ vector-boson pair at 14~TeV, in the Standard Model and beyond. We consider different 
signatures corresponding to the cleanest $H$ and $W$ decay channels, 
and discuss the potential of the high-luminosity phase of the LHC.
In particular, we investigate
  the sensitivity of the $\HWW$ production to possible anomalous Higgs couplings to  vector bosons and fermions. Since the $b$-quark initiated
partonic channel contributes significantly to this process, we find a moderate sensitivity  to both the size and sign of  an anomalous top-quark Yukawa coupling, because perturbative unitarity in the standard model implies a  destructive interference in the $b\bar b$ subprocess. We show that a combination of various signatures can reach    a 
$\sim 9$ 
standard-deviation sensitivity in the presently allowed negative region of the top-Higgs coupling, if not previously excluded.
 \end{abstract}

\maketitle

\section{Introduction}
 After the Higgs boson discovery \cite{Aad:2012tfa,Chatrchyan:2012ufa} in 2012, a  present and future major experimental task at the LHC is  to test the detailed standard model (SM) predictions for the new-particle properties and couplings to known particles. Possible non-standard Higgs couplings to both known and speculated particles are to be taken into account in  Higgs studies.
In order to characterize the Higgs boson in the most accurate way, one should then 
scrutinize not only  the main Higgs production channels, but also  the rarest  processes that can be sensitive to anomalous and/or new  kinds of  interactions. 
Here, we consider the associated production of a Higgs boson and a vector boson pair in the channel\footnote{
The  Higgs boson production in association with a pair of electroweak gauge bosons ($WW,ZZ,Z\gamma$) in $e^+e^-$ collisions
has been considered in the SM framework in 
\cite{Baillargeon:1993iw}.}
\beq
p\,p\to \HWW \, .
\label{hww}
\eeq
The cross section for the process in Eq.(\ref{hww})  is of third order in the electroweak coupling, just as the dominant Higgs boson production in $WW$ fusion. On the other hand, the phase-space factor for the production of three massive objects  depletes the total production rate at 14~TeV down to 
 about 8 fb [at leading order (LO)] \cite{Cheung:1993bm,Djouadi:2005gi}, to be compared with the  $WW/ZZ$-fusion cross section of about 4 pb. Next-to-leading order (NLO) QCD corrections enhance the $\HWW$ rates by about 50\% \cite{Mao:2009jp}. Similar considerations hold for the cross sections corresponding to the $HWZ$ and $HZZ$ final states, that are further depleted by SU(2) invariance down to about 4 fb and 2 fb at LO, respectively. The study of such relatively small cross-section 
processes  then requires the large integrated luminosities  expected in the high-luminosity phase of the LHC (HL-LHC), where one expects to collect about 3000 fb$^\mo$ of data
per experiment.

It is well-known that, in presence of anomalous Higgs couplings to vector bosons $V=W,Z$ and/or fermions $f$, 
there are processes which violate perturbative unitarity at high energies.
In particular, any  measured deviation from the SM $VVH$ and $f\bar{f}H$ couplings
results in new phenomena, since 
 further {\it unknown} degrees of freedom are necessarily required in order to recover unitarity in $V_LV_L\to V_LV_L$ \cite{Lee:1977eg} and $V_LV_L\to f\bar{f}$ scatterings \cite{Appelquist:1987cf}.

Presently,  ATLAS \cite{ATLAS:2012wma,ATLAS:2013sla} 
and CMS \cite{CMS:aya} data show a sign ambiguity in the 
Higgs couplings to fermions. The two dimensional fits of $C_V=g_{VVH}/g_{VVH}^{SM}$ and $C_f=g_{ffH}/g_{ffH}^{SM}$ (where $g_{HVV}$ and $g_{ffH}$ parametrize the Higgs couplings to  gauge bosons and fermions, respectively) are both compatible within $2\sigma$ with a SM coupling setup $C_V=C_f=1$.
On the other hand,  a non-SM fit with $C_V\simeq -C_f\simeq 1$ is not yet excluded \cite{ATLAS:2013sla}. The relative sign between the  $VVH$ and $f\bar{f}H$ couplings is  predicted by the SM (being related to the SM Higgs mechanism for the fermion mass generation), and a flipped  sign would spoil the unitarity and renormalizability of the theory. Nevertheless, there are theoretical frameworks that predict such a possibility \cite{Gabrielli:2010cw,Hedri:2013wea}.

A possible strategy to resolve the above sign degeneracy in the LHC 
data  is to look at processes where  two contributions to the scattering  amplitude, depending separately on 
the $VVH$ and $f\bar{f}H$ couplings,  
interfere. 
An example is given by the Higgs production in association with a single top in 
$pp \to t q H$, whose total cross section  gets largely enhanced by flipping the top Yukawa coupling sign in such interference contributions \cite{Tait:2000sh,Maltoni:2001hu,Barger:2009ky}.
This gives the process a considerable potential for constraining 
the negative $C_f\simeq -C_V$ coupling region 
\cite{Biswas:2012bd,Farina:2012xp}. Indeed, even the present 7+8 TeV LHC data set could be sufficient to exclude the {\it wrong}-sign Yukawa solution in
$pp \to t q H\, $ \cite{Biswas:2013xva}.
The large enhancement (by about a factor of 13 at the LHC energies) resulting from the flipped Yukawa sign in the $pp \to t q H $ cross section points 
to unitarity breaking at large energies \cite{Farina:2012xp}. Nevertheless, 
 this cross section can be reliably computed  at the LHC 
 even in the anomalous coupling region, since perturbative unitarity breaks at  energies  of the $bW\to tH$ subprocess above 10 TeV \cite{Farina:2012xp}. 
 
The larger  data sample on Higgs-boson production, expected at the LHC in forthcoming  years, will have an enormous potential to check whether the actual couplings of the newly observed particle indeed approach the corresponding SM Higgs interactions, or show some deviation from them \cite{CMS:2013xfa,ATLAS:2013hta}.

\begin{figure*}
\centering
\begin{minipage}{0.2\textwidth}
\includegraphics[width=\textwidth]{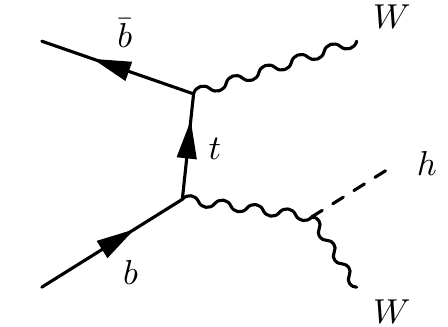}
\centering
(a)
\end{minipage}
\qquad
\begin{minipage}{0.2\textwidth}
\includegraphics[width=\textwidth]{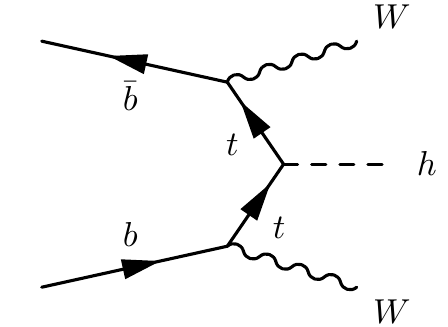}
\centering
(b)
\end{minipage}\\[1ex]
\begin{minipage}{0.2\textwidth}
\includegraphics[width=\textwidth]{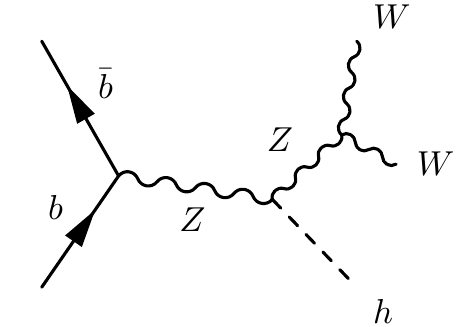}
\centering
(c)
\end{minipage}
\qquad
\begin{minipage}{0.2\textwidth}
\includegraphics[width=\textwidth]{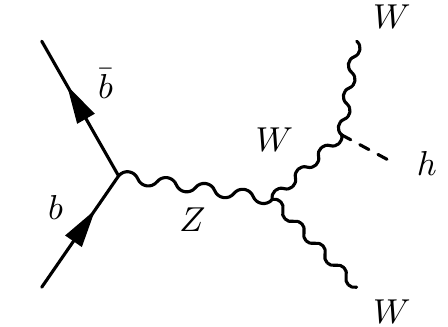}
\centering
(d)
\end{minipage}
\caption{Classes of Feynman diagrams for $b\bar{b}\to \HWW$.}
\label{fig:diagrams}
\end{figure*}

In the present analysis, we aim also to analyze what the study of $pp\to \HWW \,$
can add to the potential of other Higgs production processes characterized
by higher cross section.
 This is motivated by the fact that, in $pp\to \HWW$,
 the partonic contribution arising from the $b$-quark scattering  $b\bar{b}\to \HWW\,$  
 (Figure~\ref{fig:diagrams}) provides another example of process  sensitive to the 
top Yukawa sign (and magnitude) through the 
 interference between diagrams where the Higgs boson is radiated by a $W/Z$ boson and those ones where it is emitted by an internal top-quark 
 line. Even in this case, anomalous Higgs couplings will induce perturbative unitarity violations. Nevertheless, the possible impact of such violations on the total cross section will be diluted by the dominant light-quark scattering contribution to the $pp\to \HWW\;$ cross section, which is mostly insensitive to the Higgs Yukawa couplings.
 In the following, we will  discuss the $pp\to \HWW\,$ rate sensitivity to both Higgs Yukawa and gauge couplings, and analyze 
  the corresponding unitarity bounds in  presence of anomalous couplings.

 The paper is organized as follows. In Section II, within the SM framework, we  evaluate the $pp\to \HWW$ total cross section for different c.m energies, and compare it to the cross sections for other multi-boson final states.
In Section III, we  discuss signal versus background
 expectations at the HL-LHC,  for the most robust  
 $\HWW$ signatures (\ie multi-leptons final states, and   
di-photon resonances).
Then, in Section IV, we discuss the sensitivity of the $\HWW$ production to anomalous Higgs couplings to fermions and vector bosons. In Section V, we sum up and give our conclusions.

\begin{table}
\begin{center}
\begin{tabular}{l|r|r|r|r|r|r}

& 14 TeV & 33 TeV & 40 TeV & 60 TeV & 80 TeV & 100 TeV \\ \hline

 $\HWW$ & 8.4 & 29 & 38 & 65 & 94 & 124 \\ \hline
 
 $HWZ$ & 3.8 & 14 & 18 & 31 & 44 & 58 \\
 
 $HZZ$ & 2.1 & 7.4 & 9.6 & 16 & 24 & 31 \\

$HHW$ & 0.43 & 1.6 & 2.1 & 3.6 & 5.2 & 7.0 \\

$HHZ$ & 0.27 & 1.0 & 1.3 & 2.2 & 3.3 & 4.4 \\ \hline

$HH$ & 33.8 & 207 & 298 & 609 & 980 & 1420
\end{tabular}
\end{center}
\caption{LO electroweak tri-boson cross sections (including either one or two Higgs bosons in the final state), in $pp$ collisions (in fb) for $m_H=125$ GeV, at different c.m.  energies, and, 
for comparison, the NLO cross section for $gg \to HH$.  }
\label{tab:rare_processes}
\end{table}

\section{tri-boson cross sections}

In order to provide a context for our study, we start by overviewing the tri-boson 
electroweak final states that involve at least one Higgs
boson for the LHC energies and beyond.  In particular, we compare
the $pp \to \HWW$ cross section to the ones for other
 tri-boson final states, including either one or two Higgs bosons, at different collision c.m. energies that could be of interest at future $pp$ colliders \cite{HiggsEuStrategy}. The $HH$ production cross sections are also presented here for
comparative purposes. We postpone to the next section a detailed study of the cleanest $\HWW$ production signatures versus the most relevant backgrounds, and a discussion of the potential of the 
HL-LHC to observe the $\HWW$ process with an integrated luminosity of 3000 fb$^\mo$.

In Table \ref{tab:rare_processes} we present the  total
cross sections for  $\HWW$, $HWZ$, $HZZ$, $HHW$, $HHZ$, and $HH$ production
in proton-proton collisions for the LHC design energy of 14 TeV, and at possible future hadron colliders. From now on, we will assume $m_H=125$ GeV.

The LO cross sections 
in Table \ref{tab:rare_processes}  have been computed
with MadGraph5  \cite{Alwall:2011uj}, by using the CTEQ6L1 parton distribution functions  (PDF's) \cite{Pumplin:2002vw}. The $HHW$ and $HHZ$ cross sections have been calculated 
by retaining only the tree-level contribution of vector boson 
fusion (VBF)  from quarks initiated processes, and by neglecting the
next-to-leading contribution arising from $W/Z$ radiation by a $HH$ pair produced  via gluon-gluon fusion.
The dependence on the renormalization and factorization scales has been tested by varying the scale from a central 
value $\mu_0 = 265 \mbox{\rm GeV} \approx 2 M_W + M_H$ to $2 \,\mu_0$ and $\mu_0/2$. 
The corresponding scale uncertainty has been found  in the range $1\%-2\%$. 
For comparison, we also include in Table \ref{tab:rare_processes}
 the NLO gluon fusion cross section for $HH$ production  \cite{HiggsEuStrategy}.

The $\HWW$ production (from now on labeled just as $HWW$) turns out to have  the largest cross section  among all tri-boson channels involving Higgs bosons in the final state. Its production rate is almost a factor of 4, or 11, smaller than 
the double Higgs production at 14 TeV, or 100 TeV, respectively.
Notice that the $HH$ cross section increases with energy faster than 
all tri-boson cross sections, as the latter acquire almost a common rescaling factor while growing with energy. This behavior reflects the different evolution in energy of the gluon PDF (that mainly influences the $HH$ production) versus the quark 
PDFs, which give the dominant contribution to the 
tri-boson cross sections.

\section{$HWW$ signals and backgrounds}

In this section, we  detail our analysis of 
signatures and corresponding backgrounds for the cleanest $HWW$ decay channels.
Note that the present study partially overlaps with the analysis of the $HH\to H WW^*$
final state mediated by two Higgs-boson production \cite{Papaefstathiou:2012qe}, which has a slightly larger cross section ($\sigma_{HH}\times 2\, BR(H\to WW^*)\sim 16$ fb at 14 TeV), but differs in the presence of one ``less characterizing" {\it off-shell} $W$ in the final state.

Table \ref{Table:DecayModes} shows a list of the most relevant final states arising from the $HWW$ system decays, as well as the corresponding event numbers
at 14 TeV for 3000 fb$^{\!\scriptscriptstyle{-1}}$ (before applying any kinematical cut). One can see how multi-lepton and two-photon final states (that are the most robust against background) are in general characterized by lower rates.
\begin{table}[htbp]
\begin{tabular}{|c|c|c|c|l|}
\hline
$H\to$ & final state & BR & ev/3\,ab$^{\!\scriptscriptstyle{-1}}$ & \multicolumn{1}{c|}{signature} \\ \hline \hline
$b \bar{b}$ &  & 61\% & 16800 &  \\ \hline \hline
 & $b \bar{b}$ $\ell\nu$ $\ell\nu$ & 2.9\% & 815 & $2 b$  $ 2 \ell$ $  \met$ \\ \hline
 & $b \bar{b}$ $\ell\nu$ $jj$ & 18\% & 4960 &  $2b$   $2j$   $\ell$  $\met$ \\ \hline
 & $b \bar{b}$ $jj$ $jj$ & 27\% & 7560 &  $2b$   $4j$ \\ \hline\hline
WW* &  & 20\% & 5580 &  \\ \hline\hline
 & $\ell\nu$ $\ell\nu$ $\ell\nu$ $\ell\nu$ & 0.047\% & 13 &  $4\ell$  $\met$ \\ \hline
 & $\ell\nu$ $\ell\nu$ $\ell\nu$ $jj$ & 0.58\% & 159 &  $2j$   $3\ell$  $\met$ \\ \hline
 & $\ell\nu$ $\ell\nu$ $jj$ $jj$ & 2.6\% & 727 &  $4j$  $2\ell$  $\met$ \\ \hline
 & $\ell\nu$ $jj$ $jj$ $jj$ & 5.3\% & 1480 &  $6j$   $\ell$  $\met$ \\ \hline
 & $jj$ $jj$ $jj$ $jj$ & 4.1\% & 1120 &  $8j$ \\ \hline\hline
$\tau^+ \tau^-$ &  & 6.2\% & 1710 &  \\ \hline\hline
 & $\ell\nu \bar{\nu}$ $\ell\nu \bar{\nu}$ $\ell\nu$ $\ell\nu$ & 0.033\% & 9 &  $4\ell$  $\met$ \\ \hline
 & $\ell\nu \bar{\nu}$ $\ell\nu \bar{\nu}$ $\ell\nu$ $jj$ & 0.20\% & 55 &  $2j$   $3\ell$  $\met$ \\ \hline
 & $\ell\nu \bar{\nu}$ $\ell\nu \bar{\nu}$ $jj$ $jj$ & 0.30\% & 84 &  $4j$   $2\ell$  $\met$ \\ \hline
 & $\ell\nu \bar{\nu}$ $\tau_{\rm had}$ $\ell\nu$ $\ell\nu$ & 0.13\% & 37 & $\tau_{\rm had}$   $3\ell$  $\met$ \\ \hline
 & $\ell\nu \bar{\nu}$ $\tau_{\rm had}$ $\ell\nu$ $jj$ & 0.81\% & 223 &  $2j$  $\tau_{\rm had}$   $2\ell$  $\met$ \\ \hline
 & $\ell\nu \bar{\nu}$ $\tau_{\rm had}$ $jj$ $jj$ & 1.2\% & 340 &  $4j$  $\tau_{\rm had}$   $\ell$  $\met$ \\ \hline
 & $\tau_{\rm had}$ $\tau_{\rm had}$ $\ell\nu$ $\ell\nu$  & 0.13\% & 37 &  $2\tau_{\rm had}$   $2\ell$  $\met$ \\ \hline
 & $\tau_{\rm had}$ $\tau_{\rm had}$ $\ell\nu$ $jj$  & 0.82\% & 226 &  $2\tau_{\rm had}$   $2j$   $\ell$  $\met$ \\ \hline
 & $\tau_{\rm had}$ $\tau_{\rm had}$ $jj$ $jj$ & 1.2\% & 345 &  $2\tau_{\rm had}$   $4j$ \\ \hline\hline
ZZ* & \multicolumn{1}{l|}{} & 2.5\% & 690 &  \\ \hline\hline
\multicolumn{1}{|l|}{} & $\ell\ell$ $\ell\ell$ $\ell\nu$ $\ell\nu$ & 0.001\% & 0 & $6\ell$  $\met$ \\ \hline
\multicolumn{1}{|l|}{} & $\ell\ell$ $\ell\ell$ $\ell\nu$ $jj$ & 0.003\% & 1 &  $2j$   $5\ell$  $\met$ \\ \hline
\multicolumn{1}{|l|}{} & $\ell\ell$ $\ell\ell$ $jj$ $jj$ & 0.005\% & 1 &  $4j$   $4\ell$ \\ \hline
\multicolumn{1}{|l|}{} & $\ell\ell$ $jj$ $\ell\nu$ $\ell\nu$ & 0.006\% & 2 &  $2j$   $4\ell$  $\met$ \\ \hline
\multicolumn{1}{|l|}{} & $\ell\ell$ $jj$ $\ell\nu$ $jj$  & 0.017\% & 5 &  $4j$   $3\ell$  $\met$ \\ \hline
\multicolumn{1}{|l|}{} & $\ell\ell$ $jj$ $jj$ $jj$  & 0.053\% & 15 &  $6j$   $2\ell$ \\ \hline
\multicolumn{1}{|l|}{} & $jj$ $jj$ $\ell\nu$ $\ell\nu$ & 0.059\% & 16 &  $4j$   $2\ell$  $\met$ \\ \hline
\multicolumn{1}{|l|}{} & $jj$ $jj$ $\ell\nu$ $jj$  & 0.36\% & 100 &  $6j$  $\ell$  $\met$ \\ \hline
\multicolumn{1}{|l|}{} & $jj$ $jj$ $jj$ $jj$  & 0.55\% & 152 & $8j$ \\ \hline\hline
$\gamma \gamma$ & \multicolumn{1}{l|}{} & 0.22\% & 61 &  \\ \hline\hline
\multicolumn{1}{|l|}{} & $\gamma \gamma$ $\ell\nu$ $\ell\nu$ & 0.011\% & 3 &  $2\gamma$  $2\ell$  $\met$ \\ \hline
\multicolumn{1}{|l|}{} & $\gamma \gamma$ $\ell\nu$ $jj$ & 0.065\% & 18 & $2\gamma$  $2j$  $\ell$  $\met$ \\ \hline
\multicolumn{1}{|l|}{} & $\gamma \gamma$ $jj$ $jj$ & 0.099\% & 27 & $2\gamma$  $4j$ \\ \hline
\end{tabular}
\caption{List of most relevant final states arising from the $HWW$ system decays, and the corresponding event numbers at 14 TeV for 3000 fb$^{\!\scriptscriptstyle{-1}}$, before applying any kinematical cut.}
\label{Table:DecayModes}
\end{table}


In the following, both signal and background event numbers have been worked out by using MadGraph 5 \cite{Alwall:2011uj}, 
interfaced with Pythia 6.4  \cite{Sjostrand:2006za} for decays  with large particle multiplicities. 
All event samples have been analyzed at parton level. 
The following set of basic kinematical cuts has been universally applied in this paper:
\begin{itemize}
\item for final state leptons ($e$, $\mu$) and photons, we require a pseudo-rapidity cut $|\eta|<2.5$, and a transverse momentum cut $p_T > 10$ GeV;
\item for final state quark and gluon jets, we impose $|\eta|<2.5$ and $p_T > 20$ GeV. We disregard
forward jets with $|\eta|>2.5$ to ensure that $b$-jets can be more reliably identified, 
$b$-tagging algorithms being more efficient in the central part of the detector. We assumed a $b$-jet detection efficiency of 70\%;
\item for each pair of visible objects ({\it i,j}\,), we require an isolation cut  $\Delta R_{ij} > 0.4$, where
$\Delta R_{ij}=\sqrt{\eta^2_{ij}+\phi^2_{ij}}$, and $\eta_{ij} (\phi_{ij})$ is their rapidity (azimuthal)  separation.
\end{itemize}
In order to investigate hadronic tau decays, we have modified the Tauola code in MadGraph 
to assign a unique particle identifier to the hadronic tau decay products, $\tau_{\rm had}$. We then applied to $\tau_{\rm had}$ the same
set of cuts as adopted for quark and gluon jets.

In our analysis, we do not include decay channels into N jets plus two opposite-sign leptons, or one single lepton, or no leptons, which are
dominated by QCD backgrounds such as top-pair production. This  excludes  
the highest-rate (but challenging) channels with Higgs into $b \bar{b}$  \footnote{The $WW b\bar{b}$ channel has been proposed as a signal channel for $HH$ production \cite{Papaefstathiou:2012qe}. In this case the presence of two on-shell Higgs bosons (implying at least one very off-shell $W$)  provides additional kinematic constraints to reject the top background.}.
We also disregard the Higgs decay into   $ZZ^*$ that generates too few events into the most-robust semi-leptonic/all-leptonic final states.

\subsection{Four Lepton Final States}

 $HWW$ final states can go to four charged leptons  either via $H \to WW^{*}\to \ell\nu \ell\nu$, or via $H \to \tau^+ \tau^-\to \ell\nu \nu\, \ell\nu\nu$, with the accompanying on-shell $W$ pair also decaying leptonically, $WW\to \ell\nu \ell\nu$.  In either cases, there is significant missing energy due to either 4 or 6 neutrinos.
The main irreducible backgrounds for these  channels are
\begin{itemize}
\item EW continuum production of four $W$'s,  all  decaying leptonically;
\item $WWZ$ with $Z$ decaying into either $ee,\mu\mu$, or  two leptonically decaying $\tau$'s. The former channel has higher rates, but can be tamed by cutting away the $Z$ mass region from the lepton pair invariant mass distribution (indeed we do not include any
background presenting a $Z\to ee,\mu\mu$ resonance throughout the present analysis). The leptonic $Z$ decay via $\tau^+ \tau^-$ (that we include) has lower rates, but is less characterized, and in general more overlapped with the Higgs signal into leptons;
\item $ZZ$ pairs with the $Z$'s decaying into $e,\mu$ leptons or leptonically decaying taus; again the decays $Z\rightarrow \ell^+\ell^-$ can be cut away by reconstructing the $Z$ resonance;
\item $ZH$ with  $Z$ decaying into $e,\mu$ leptons or leptonically decaying taus, and  $H$ decaying to leptons through $WW^*$ or  taus pairs;
\item $HH$ with both $H$'s going into $WW^*$, followed by leptonic decays of the $W$ bosons.
\end{itemize}

The corresponding signal and background rates are shown in Table \ref{cutflow_4leptons}. 
We find that the
most dangerous irreducible background, after our basic kinematic cuts, comes from $ZZ$ production,
with both $Z$'s decaying into $\tau$'s. In order to reduce this background, we cut on the scalar sum of the missing energy and the transverse momentum of the four leptons,
$\sum_i p_T^{\ell_i}+\met$. Indeed, leptons from the
indirect decays via $\tau \to \ell \nu\nu$ are typically produced with lower transverse energy. The total
missing energy in the event is also  lower for  indirect decays, since the eight neutrinos in each event are on average emitted in random directions 
on the transverse plane, and their momenta partially cancel each other out. 
We find that, by applying a lower cut of 200 GeV on  
 $\sum_i p_T^{\ell_i}+\met$, the $ZZ$ background can be reduced by about a factor of 10, while the signal falls  by 25\% only.

After all the cuts described above,  the signal to background ratio is close to 0.32, and the corresponding significance ($S/\sqrt{S+B}$ in unity of standard
deviation $\sd$) of the four-lepton channel 
for a dataset of 3000 fb$^\mo$ is about $1.3\, \sd$.

\begin{table}
\begin{center}
\begin{tabular}{l|r|r}
& $\;$ basic cuts $\;$ & $\; \sum p_T + \met$ \rule[-1ex]{0ex}{2ex} \\ \hline
$HWW$ signal: \rule{0ex}{3ex} & & \\
$\quad$ via $H \to WW^*$ & 2.1 & 1.6 \\
$\quad$ via $H \to \tau\tau$ & 1.0 & 0.8 \\ \hline
$ZZ \to 4 \,\tau \to 4\, \ell +\nu's$ \rule{0ex}{3ex} & 17.9 & 1.7 \\
$WWZ$ & 3.3 & 2.5 \\
$4\,W$ & 0.7 & 0.7 \\
$ZH, \; H \to WW^*$ & 0.7 & 0.2 \\
$ZH, \; H \to \tau\tau$ & 1.6 & 0.4 \\
$HH\to WW^*WW^*$ & 3.1 & 2.1 \\ \hline \hline
total signal & 3.1 & 2.4 \\
total background & 27.3 & 7.6 \\
\end{tabular}
\end{center}
\caption{Signal and background cut flow for $4 \,\ell + \met$ final states. All cross sections are in ab. In the third column, $\sum p_T + \met$ labels the cut on the scalar sum of  missing energy and  four-lepton transverse momenta, $\sum_i p_T^{\ell_i}+\met>200$ GeV.  }
\label{cutflow_4leptons}
\end{table}

\subsection{Hadronic $W$ + 3 leptons}

We will now investigate the channel with one hadronic $W$ decay ($W_{\rm had}$) plus three charged leptons in the final 
state. As in the previous channel, this signature can arise from the $HWW$  state in connection to two different Higgs decays. 
For the Higgs decaying into $WW^*$, compared to the four-lepton final state, the rate is 
 enhanced by two effects:
\begin{itemize}
\item the $W$ branching ratio (BR) into hadrons is more than a factor 3 larger than the one into $e \nu+\mu \nu$;
\item any of the three on-shell $W$'s can decay hadronically, giving a further factor of 3 from combinatorial enhancement.
\end{itemize}
On the other hand, when the signature arises from $H \to \tau \tau \to$ leptons, 
the $W$ BR is again increased by a factor of 3, while  the combinatorial factor is only 2.
Altogether, after applying kinematical cuts, about 80\% of the signal events
originates from the $H \to WW^*$ decay mode (cf. Table~\ref{cutflow_3leptons}).

We will now discuss the (mostly irreducible) backgrounds that can lead to the $W_{\rm had}$ + 3 $\ell$ + $\met$ final state, in order of relevance:
\begin{itemize}
\item   2 jets + $WZ$, where the $Z$ decays via $\tau$'s to leptons. The total rate for this background after basic kinematic cuts is larger 
than 1 pb. On the other hand, by requiring the jet pair to 
reconstruct the $W$ mass, it falls down by a factor 20;
\item  $t \bar{t} W$  production, 
where the two $b$-jets from top decays are mis-tagged as light jets, and reconstruct 
 the $W$ mass.  By demanding the jets to reconstruct the $W$ 
mass within 5 GeV,  the $t\bar{t}W$ background has been reduced by a factor $\sim 24$
(cf. Table~\ref{cutflow_3leptons}). Similarly, the $t \bar{t} Z$ production, for $Z\to \ell\ell$, can contribute to the background whenever one of the charged leptons from the $Z$ decay falls outside the experimental acceptance (this actually occurs in about 1/6 of the events);
\item further  QCD backgrounds originate from $jjWWW$ and 
$jjWH$, but can similarly be reduced by requiring the jet-pair invariant mass to reconstruct  $M_W$;
\item purely electroweak backgrounds, which mainly originate from 
$4W$ and $WWZ$ production. For $WWZ$, we assume that the $Z$ decays via $\tau$'s to leptons;
\item $HH\to WW^*WW^*$ production. Note that $HH$ is more affected by previous cuts, since two out of four $W$'s are off-shell, and, for $W$ hadronic decays,  do not reconstruct $M_W$, while,  for $W$ leptonic decays, give reduced transverse momenta.
\end{itemize}

Then, in general, in addition to our basic kinematic cuts, we demand the two jets to reconstruct
$M_W$ within a mass window of $\pm$ 5 GeV.
The effect of the above cuts on signal and background is shown in Table \ref{cutflow_3leptons}.
With 3000 fb$^\mo$, we expect 69 signal and 477 background 
events. The $S/B$ ratio is 0.145 and the corresponding significance is $3.0\, \sd$.

\begin{table}
\begin{center}
\begin{tabular}{l|r|r} 
& $\; \;$ basic cuts $\; \;$ & $\; \; m(W) \; \;$ \\ \hline
$HWW$ signal: \rule{0ex}{3ex} & & \\
$\quad$ via $H\to WW^*$   & 22.3 & 18.8 \\
$\quad$ via $H\to \tau\tau$ \rule[-1ex]{0ex}{2ex} & 4.3 & 4.3 \\ \hline
$WWZ$ \rule{0ex}{3ex} & 17.7 & 17.7 \\
$4W$ & 7.0 & 7.0 \\
$jjWWW$ & 740 & 29.4 \\
$jjWZ$ & 1540 & 49.9 \\
$jjWH$, $H \to WW^*$ & 169 & 9.7 \\
$jjWH$, $H \to \tau\tau$ & 82.2 & 4.4 \\
$t\bar{t}W$ & 825 & 34.9 \\ 
$t\bar{t}Z$ & 11.7 & 0.5 \\ 
$HH \to WW^*WW^*$ & 10.3 & 5.4 \\ \hline \hline
total signal & 26.6 & 23.1 \\
total background & 3400 & 159 

\end{tabular}
\end{center}
\caption{Signal and background cut flow in the $3 \ell + W_{\rm had} $ final states. All cross sections are in ab. The cut labeled $m(W)$ requires  the invariant mass of the two jets to satisfy $75.4 \; {\rm GeV} < m^{jj} < 85.4 \; {\rm GeV}$.}
\label{cutflow_3leptons}
\end{table}

\subsection{Higgs decay into diphotons}

We now examine the final states where the Higgs decays to two photons. The resonant
$\gamma\gamma + WW$ signal is very clear, and the backgrounds are in general small, but the signal is penalized by the small Higgs BR to photons. The cleanest signature to look for would obviously be the full leptonic $WW$ final state, but the corresponding rate is highly suppressed, giving a total of about 3 events with 3000 fb$^\mo$ (cf. Table~\ref{Table:DecayModes}). Thus we concentrate on the larger-rate semi-leptonic $WW$ channel, resulting in the final state $jj\ell\nu\gamma\gamma$. The main irreducible backgrounds are
\begin{itemize}
\item $j j W \gamma \gamma$, with  the $W$ decaying into leptons, where the jets reconstruct the $W$ mass, and the photons reconstruct  the Higgs mass;
\item $j j W H$, that is $WH$ associated production with two extra
 jets faking a hadronic $W$ decay;
 \item $WW\gamma\gamma$, with one  $W$ decaying leptonically, and the other one hadronically, and two radiated photons that reconstruct the Higgs system;
\item $H H$, with one of the Higgs bosons decaying into two photons and the other one into a semi-leptonic $W$ pair. One of the $W$'s from the Higgs decay being off-shell,  this background will be reduced by a proper cut around $M_W$ on  the hadronic $W^*$.
\end{itemize}

The two-jet and two-photon invariant masses are then required  to be within the ranges $M_W\pm5$ GeV, and  $m_H\pm2$ GeV, respectively. The main backgrounds contain radiated jets faking the $W$, and/or radiated photons faking the Higgs. The $p_T$ spectrum of the radiated objects is softer than the corresponding spectrum for the decay
products of a real $W$ and Higgs, so we require additional cuts on the scalar $p_T$ sum of the two jets and the two photons, respectively, as $p_T^{j_1}+p_T^{j_2} > 70$ GeV and $p_T^{\gamma_1}+p_T^{\gamma_2}>100$ GeV. The effect of the above cuts on signal and background is shown in Table \ref{cutflow_digamma_semileptonic}. As a results, one gets a  $jj\ell\nu\gamma\gamma$ signal to background ratio of about 2/9, and a significance $\simeq 1.0\, \sd$ for a 
dataset of 3000 ${\rm fb}^\mo$.

\begin{table}
\begin{center}
\begin{tabular}{l|r|r|r|r|r} 
& $\;$ basic cuts $\;$ & $\; m(H) \;$ & $\; \Sigma p_T^\gamma \;$ & $\; m(W) \;$ & $\; \Sigma p_T^j \;$ \rule[-1ex]{0ex}{2ex} \\ \hline
$HWW$ signal \rule{0ex}{3ex} & 2.5 & 2.5 & 2.3 & 2.0 & 1.95 \\ \hline
$jjW\gamma\gamma$ \rule{0ex}{3ex} & 18400 & 144 & 105 & 4.3 & 4.1 \\
$jjWH$ \rule{0ex}{3ex} & 61 & 61 & 55 & 3.1 & 2.5 \\
$WW\gamma\gamma$ \rule{0ex}{3ex} & 264 & 2.4 & 1.7 & 1.5 & 1.4 \\
$HH$ \rule{0ex}{3ex} & 2.0 & 2.0 & 1.8 & 1.22 & 1.18 \\ \hline \hline
total signal & 2.5 & 2.5 & 2.3 & 2.0 & 1.95  \\
total backgr. & 18700 & 209 & 164 & 10.1 & 9.2 \\ 
\end{tabular}
\end{center}
\caption{The cut-flow for the 2 jets + 1 $\ell$ + 2$\gamma$ final states. All cross sections are in ab. The labels are defined as follows: $m(H)$ stands for a cut on the $\gamma\gamma$ invariant mass, $123\, {\rm GeV} < m^{\gamma\gamma} < 127\, {\rm GeV}$, $\Sigma p_T^\gamma$ is a cut on the scalar sum of the photon transverse momenta, $p_T^{\gamma_1}+p_T^{\gamma_2}>100$ GeV, $m(W)$ is a cut on the jet-pair invariant mass, $75.4 {\rm GeV} < m^{jj} < 85.4 {\rm GeV}$, and $\Sigma p_T^j$ is a cut on the scalar sum of the jet transverse momenta, $p_T^{j_1}+p_T^{j_2} > 70$ GeV.}
\label{cutflow_digamma_semileptonic}
 \end{table}

\subsection{Same sign leptons from $H \to \tau \tau$}

We now discuss the  $ \ell\nu j j \tau \tau$ final state, arising from  semi-leptonic  $WW$ decays, and Higgs decaying to $\tau$ pairs. We consider the case where one of the taus decays leptonically, 
and the other hadronically, and demand   two same-sign leptons (one from a $W$ and one from a $\tau$). The signature is therefore two jets, two 
same-sign leptons, and one hadronic tau ($\tau_{\rm had}$). We select the latter  channel since  
same-sign lepton events are very much suppressed in the SM, so that
even a small number of signal events could lead to an observation.

The main irreducible backgrounds for this signature are 
\begin{itemize}
\item $W W Z$, with $Z \to \tau \tau$. Because of the missing energy from neutrinos in tau decays, the mass of the particle decaying to  taus can not be reconstructed accurately, and the $Z$ and
Higgs signals will be  in general quite overlapped;
\item $j j W H$, \ie $WH$ associated production with two extra
 jets, and $H \to \tau \tau$;
 \item $j j W Z$, with  $Z \to \tau \tau$. This is the main background for this channel because of the large production cross section. Again, the $Z$ and Higgs decay products 
via taus will be in general  quite overlapped.
\end{itemize}

In addition to our basic kinematic cuts, we again demand 
the two-jet  mass  to be within the range $M_W\pm5$ GeV,
 and  assume 100\% efficiency for hadronic tau identification. The effect of kinematical  cuts on the signal and background is shown in Table \ref{cutflow_samesign}. After all cuts,  the same-sign lepton  signal to background ratio for $ \ell\nu j j \tau \tau$ 
is about 0.034, and the significance for a data set of 3000 ${\rm fb}^\mo$ is $0.74\, \sd$.

\begin{table}
\begin{center}
\begin{tabular}{l|r|r} 
& $\;$ basic cuts $\;$ & $\; m(W) \;$ \rule[-1ex]{0ex}{2ex} \\ \hline
$HWW$ signal \rule{0ex}{3ex} & 5.6 & 5.6 \\ \hline
$WWZ$ \rule{0ex}{3ex} & 27.6 & 27.6 \\
$jjWH$ \rule{0ex}{3ex} & 106 & 7.7 \\
$jjWZ$ \rule{0ex}{3ex} & 2820 & 129 \\ \hline \hline
total signal & 5.6 & 5.6 \\
total background & 2950 & 164 \\ 
\end{tabular}
\end{center}
\caption{The cut-flow for the 2 jets + $\ell^{\pm}\ell^{\pm}$ + $\tau_{\rm had}$  final states. All cross sections are in ab. The  label $m(W)$ stands for a cut on the jet-pair invariant mass, $75.4 \,{\rm GeV} < m^{jj} < 85.4\, {\rm GeV}$.}
\label{cutflow_samesign}
\end{table}

\subsection{Same sign leptons from $H\to W^+W^-$}

The same sign dilepton signal can also arise from $W^+W^-H\to W^+W^-W^+W^-$, where the positively charged $W$'s decay to leptons and the negatively charged $W$'s into hadrons or vice versa. In this case the final state consists of two hadronic $W$ systems, two same sign leptons and missing energy. The most relevant backgrounds are:

\begin{itemize}
\item $4jW^\pm W^\pm$, where the $W$'s decay into leptons, and the four jets fake the two hadronic $W$'s. Because of the valence-quark charge distribution, the cross section of the positively charged $W$ pair production is about three times as large as the negative pair one.
\item $jjW^\pm W^\pm W^\mp $, where the same sign $W$'s decay into leptons, and the opposite sign $W$ decays into hadrons, and the two jets reconstruct the remaining hadronic $W$.
\item $t\bar{t}W^\pm \to b\bar{b}W^+W^-W^\pm$, where the same sign $W$'s decay into leptons and the opposite sign $W$ decays into hadrons, and the two $b$-jets fake the hadronic $W$.
\item $tW^+W^-j \to W^+bW^+W^-j$, or the charge conjugate process, where the same sign $W$'s decay into leptons and the opposite sign $W$ into hadrons, and the $b$-jet plus the light jet fake the hadronic $W$.
\item $W^+W^+W^-W^-$, that is electroweak production of four $W$ bosons, with  hadronic (leptonic) decays of the positively (negatively) charged $W$'s, or 
viceversa.

\end{itemize}

To extract the signal from the background we require the four jets in the final state to combine into two pairs with invariant mass within $\pm5$ GeV around $M_W$. To reduce the background from the $t\bar{t}W$ production, events with $b$-tagged jets are vetoed, (assuming a 70\% tagging efficiency). The resulting signal and background rates are shown in table \ref{cutflow_samesign_W_neg} for negatively charged leptons and in table \ref{cutflow_samesign_W_pos} for positively charged leptons. Because of the valence-quark charge distribution, the background cross sections are generally smaller for the negatively charged lepton pair, and hence this channel is more significant. After cuts, the signal to background ratio is 0.087 (0.042) for negative (positive) sign leptons, and the combined significance for $3000\, {\rm fb}^\mo$ is $0.98\,\sigma$.

\begin{table}
\begin{center}
\begin{tabular}{l|r|r} 
& $\;$ basic cuts $\;$ & $\; m(W) \;$ \rule[-1ex]{0ex}{2ex} \\ \hline
$HWW$ signal \rule{0ex}{3ex} & 4.3 & 2.7 \\ \hline
$4jWW$ \rule{0ex}{3ex} & 828 & 2.5 \\
$2jWWW$ \rule{0ex}{3ex} & 406 & 18.2 \\
$t\bar{t}W$ \rule{0ex}{3ex} & 138 & 7.7 \\
$tWWj$ \rule{0ex}{3ex} & 112 & 2.5 \\
$WWWW$ \rule{0ex}{3ex} & 0.3 & 0.3 \\ \hline \hline
total signal & 4.3 & 2.7 \\
total background & 1480 & 31.2 \\ 
\end{tabular}
\end{center}
\caption{The cut-flow for 4 jets + $\ell^-\ell^-$  final state. All cross sections are in ab. The  label $m(W)$ stands for the cut on the invariant mass of the two jet-pairs, $75.4 \,{\rm GeV} < m^{jj} < 85.4\, {\rm GeV}$. The event passes the $m(W)$-cut  for  any  possible combination of the two pair systems built from the four jets, where both   pairs pass the cut.}
\label{cutflow_samesign_W_neg}
\end{table}

\begin{table}
\begin{center}
\begin{tabular}{l|r|r} 
& $\;$ basic cuts $\;$ & $\; m(W) \;$ \rule[-1ex]{0ex}{2ex} \\ \hline
$HWW$ signal \rule{0ex}{3ex} & 4.3 & 2.7 \\ \hline
$4jWW$ \rule{0ex}{3ex} & 2830 & 14.2 \\
$2jWWW$ \rule{0ex}{3ex} & 679 & 30.2 \\
$t\bar{t}W$ \rule{0ex}{3ex} & 262 & 12.9 \\
$tWWj$ \rule{0ex}{3ex} & 249 & 6.2 \\
$WWWW$ \rule{0ex}{3ex} & 0.3 & 0.3 \\ \hline \hline
total signal & 4.3 & 2.7 \\
total background & 4020 & 63.8 \\ 
\end{tabular}
\end{center}
\caption{ Same as in Table~\ref{cutflow_samesign_W_neg}  for 4 jets + $\ell^+\ell^+$  final states.}
\label{cutflow_samesign_W_pos}
\end{table}
\subsection{Further relevant backgrounds}

The associated production of a Higgs boson and a $t\bar{t}$ pair gives a common  background for all final states investigated above, since  the final state $Ht\bar t\to HWWb\bar{b}$  can mimic the signal $HWW$ whenever the final $b$-jets are not reconstructed.

The LO cross section for the process $pp \rightarrow Ht\bar t\to HWWb\bar{b}$ is about 360 fb
at $\sqrt S=14$ TeV. If we require, in a parton-level simulation, that the $b$-jets  have transverse momentum $p_T>20$ GeV and pseudorapidity $|\eta|<4.5$ to be reconstructed at least as  further light jets, then both  $b$-jets will be reconstructed in 91.9\% of the events, and at least one $b$-jet will be reconstructed in 99.75\% of the events. Thus the $Ht\bar t$ background can be effectively suppressed down to 0.25\% of the original cross section by a veto on any additional jets with $p_T>20$ GeV and $|\eta|<4.5$. Then, the
 latter acceptance cuts reduce the cross section of the $t\bar{t}H$ background to about 0.9 fb, before applying the relevant BR's for the Higgs and $W$ bosons for each final state.  
On the other hand, the inclusion of extra QCD radiation and shower effects will in general impact the present conclusion.

A further potentially dangerous background for the 
$W_{\rm had}+3\ell$ signal is the $t\bar t j \to 2\ell 2\nu b \bar b j$
production, where a $b$-jet is mis-tagged as a light jet, and the corresponding $bj$ reconstruct
a $W_{\rm had}$, while the second $b$ is mis-tagged as a lepton \cite{Sullivan:2010jk}. Similarily for the $4\ell$ signal there is a potential background from $t\bar t \to 2\ell 2\nu b \bar b$ production, where both $b$'s are identified as leptons, although this background is suppressed by the square of the mis-tag rate. 

The impact of the latter backgrounds critically depends on the actual detector performances. 
Although backgrounds of this type, originating from mis-tags, fakes and detector effects, are likely to be relevant for the actual experimental analysis of the $HWW$ production, their detailed analysis is beyond the scope of the present work.
\subsection{Combination}
We now  combine the potential of  the six channels previously discussed, reported in 
Table  \ref{summary_Table}.
\begin{table}
\begin{center}
\begin{tabular}{l|r|r|c} 
final state $\;$ & $\;$ signal $\;$ & $\;$ backgr. $\;$ & $\; S/\sqrt{S+B}\;$ \rule[-1ex]{0ex}{2ex} \\ \hline
$4\,\ell + \met$ \rule{0ex}{3ex} & 2.4 & 7.6 & 1.3 \\
$3\,\ell+ 2j $  \rule{0ex}{3ex} & 23.1 & 159 & 3.0 \\
$1\,\ell +2j + 2\gamma $ \rule{0ex}{3ex} & 1.95 & 9.2 & 1.0 \\
$\ell^\pm \ell^\pm+2j+\tau_{\rm had}$ \rule{0ex}{3ex} & 5.6 & 164 & 0.74 \\ 
$\ell^- \ell^-+4j$ \rule{0ex}{3ex} & 2.7 & 31.2 & 0.80 \\
$\ell^+ \ell^++4j$ \rule{0ex}{3ex} & 2.7 & 63.8 & 0.57 \\ \hline \hline
total \rule{0ex}{3ex} & & & 3.6
\end{tabular}
\end{center}
\caption{Signal versus background rates (in ab) after all dedicated cuts for different final states, and the corresponding significance in unity of $\sd$ ($S/\sqrt{S+B}$)   for 3000 ${\rm fb}^\mo$. The total significance of $3.6\, \sd$  is the  sum in quadrature of all individual significances.}
\label{summary_Table}
\end{table}
Here we combine the  final rates, after the optimization procedure,  for the signal and total 
background for each final state,  and the corresponding significances.
Significances are for 3000 fb$^\mo$ of integrated luminosity.
By summing in quadrature the significances of each individual channel,
we get a total $HWW$ signal significance of $3.6\, \sd$ in the SM.
\section{Anomalous Higgs couplings}
In this section we consider the possibility that the Higgs boson has
non-SM couplings to $W,Z$ bosons and fermions.
In order to parameterize any deviation from the SM expectations, 
we introduce the set of scaling coefficients $C_{W,Z,f}$ defined as 
\bea
C_{W} &=& \frac{g_{WWH} }{g_{WWH}^{SM}}\, ,~~~C_{Z} \,=\, \frac{g_{ZZH} }{g_{ZZH}^{SM}}\, ,~~~
C_{f} \,=\, \frac{g_{ffH}}{g_{ffH}^{SM}}\, ,
\label{CX}
\eea
where $g^{SM}_{WWH}$, $g^{SM}_{ZZH}$ and $g^{SM}_{ffH}$, stand for the corresponding SM couplings. The $C_Z$ and $C_W$ parameters are constrained to be positive, while $C_f$ can still assume negative values \cite{Hedri:2013wea}.

As discussed in Section I, anomalous Higgs 
couplings to SM weak gauge bosons and fermions can induce a violation of  perturbative unitarity at some energy scale, which depends on the particular process considered. Perturbative  unitarity can then be recovered by introducing new
weakly coupled degrees of freedom with a mass 
spectrum at, or below, the unitarity breaking scale.
In case no new elementary particle appear in the spectrum, the energy scale associated to the breaking of perturbative unitarity should be interpreted as the scale where  interactions of the  Higgs boson
and  longitudinal modes of vector gauge bosons become strong 
\cite{Chanowitz:1986hu,Chanowitz:1987vj,Dobado:1995qy}. 
Unitarity is then expected to be recovered in a non-perturbative regime, by the exchange of strongly-interacting composite resonances.

In case the Higgs couplings  are modified without extending the SM content  
 below the scale of the unitarity violation, total cross sections might increase with energy faster than the corresponding SM ones. 
A relevant example is provided by the single top production  in association with a  Higgs boson mediated by the sub-process  
$W b\to H t$ in $pp$ collisions  \cite{Biswas:2012bd,Farina:2012xp,Biswas:2013xva}. Its cross section is very sensitive not only to the magnitude of the
 ratio $C_t/C_W$, but
also to its sign, because of the strong destructive interference  
between the  diagrams involving the Higgs coupling to the $W$ and to the top-quark in the SM.

For the $pp\to HWW$ production, the cross section 
receives the largest contribution from the $HWW$ coupling, and has
 a milder dependence on the $HZZ$ and $Htt$ couplings\footnote{
We neglect any contribution from   light-quark 
transitions to a top-quark ($d,s\to t$) via  $W$ exchange, the latter being  strongly suppressed by  off-diagonal terms of the Cabibbo-Kobayshi-Maskawa matrix.}.
In particular, the top-Yukawa coupling $g_{ttH}$ enters through the subprocess $b\bar{b}\to HWW$
(see Figure~\ref{fig:diagrams}), which  moderately contributes to the cross section with respect to the light-quark initiated sub-process 
$q\bar{q}\to HWW$.
As for the $HZZ$ coupling, it enters  only through the s-channel in all the sub-processes, and its  impact is therefore sub-dominant with respect to the  $HWW$-coupling one. 

When assuming anomalous couplings, 
the energy scale of the partonic process must be held below 
the characteristic scale of unitarity violation in order to keep the cross section within the perturbative regime. In the  $pp\to HWW$ case, this scale 
will mostly depend on the coefficients $C_{W,Z,t}$, and should
tend to infinity for $C_{W,Z,t}\to 1$, which recovers the SM case.
\\
In order to determine the $pp\to HWW$ sensitivity to anomalous $C_{W,Z,t}$ coefficients in a perturbative regime,  
the effective partonic c.m. energy of the $HWW$ system ($\lsim$  TeV at the LHC) must be kept below the energy scale of unitarity violations. To this purpose, we analyze below the relevant unitarity bounds associated to the  partonic processes contributing to   $pp\to HWW$ as a function of the anomalous Higgs couplings.

\subsection{Analytical unitarity bounds}
We now analyze the contribution to the $pp\to HWW$ cross section that comes from the $b$-quark initiated process
\begin{equation}
	b (p_b) \, \bar b(p_{\bar{b}}) \, \to\,  H(p_h)\, W^{\!\scriptscriptstyle{+}}(p_{+})\, W^{\!\scriptscriptstyle{-}}(p_{-})\,  ,
\label{bbWWH}
\end{equation}
where the quantities  in  parenthesis label the corresponding particle momenta.
A representative set of Feynman diagrams for the $b\bar{b}\to HWW$  
is given in Figure~\ref{fig:diagrams}.
This sub-process  receives a large contribution from the top-quark Yukawa coupling 
[see Figure~\ref{fig:diagrams}(b)], and  is also
sensitive to anomalous Higgs couplings in both  the $W,Z$ and top-quark sectors. 
 In the following, we will 
retain only the contribution from top-quark exchange diagrams, setting to zero
the Yukawa couplings of lighter quarks since their effect does not significantly affect the present results.

The  breaking of perturbative unitarity in  the process in Eq.~(\ref{bbWWH}) is induced by the contributions of the vector-boson longitudinal polarizations. At high energy, the corresponding polarization vectors are approximated by
\bea
\epsilon_L^{\mu} (p_{\pm}) \approx \frac{p_\pm^{\mu}}{M_W}\, .
\eea
By retaining only the contribution of $W^{\pm}$ longitudinal polarizations in the relevant amplitude (labelled as $\mathcal{M}_{LL}$), one gets  the asymptotic expression
\bea
&i\mathcal{M}_{LL}&=\, \frac{2 m_t^{2}}{v^{3}}\,\left(C_t-C_W\right)
\times\nonumber\\
&&\!\!\!\!\!\!\!\!\!\!\!\!\bar v_{\bar b}\left(
\frac{\slp_{-}}{(p_{b} - p_{-})^{2} - m_t^{2}} - \frac{\slp_{+}}{(p_{\bar b} - p_{+})^{2}- m_t^{2}} \right)P_L \,u_{b}+\nonumber\\
&&\!\!\!\!\!\!\!\!\!\!\!\!\!\frac{2 M_Z^{2}}{v^{3}}\left(C_{Z} - C_{W}\right)\bar v_{\bar b}\left(\frac{\slp_{+} - \slp_{-}}{s}\right)(2q_b s_w^{2} + P_L) u_b\, ,
\label{MLL}
\eea
where  $u_{b}$ and $\bar v_{\bar b}$  are the spinors of the $b$ and $\bar b$ quarks, respectively, $v$ is the Higgs vacuum expectation value, $q_b= -1/3$ is the bottom-quark electromagnetic charge, and $P_L=(1-\gamma_5)/2$ is the  left-handed chirality projector. $M_Z$ and $m_t$ are the $Z$ and top-quark mass, respectively, while the $b$-quark is assumed massless.

In the high energy limit the kinematics is simplified by treating all the external particles as massless. Under this assumption, the phase space of the final state can be parametrized by the following dimensionless variables evaluated in the c.m. frame
\bea
\nonumber
 (p_{+} + p_{-})^{2} &\simeq& s - 2 \sqrt{s}E_{H} \equiv s \,y_{H}\,,
\\ \nonumber
(p_H + p_{+})^{2} &\simeq& s - 2 \sqrt{s}E_{-} \equiv s \,y_{-}\,,
\\
 (p_H + p_{-})^{2} &\simeq& s - 2 \sqrt{s}E_{+} \equiv s \,y_{+}\, ,
\label{kinem}
\eea
(with  $y_{+} + y_{-} + y_{H} = 1$) 
and three angular variables. $E_i$ denotes the energy of the  particle $i$. The r.h.s of Eq.(\ref{kinem}) is explicitly evaluated by assuming the massless approximation.

The differential phase space, $\td \Phi_{3}$, is then expressed as
\begin{align}
	\td \Phi_{3} = \frac{s}{32 (2\pi)^{3}} \delta(1 - y_{+} - y_{-} - y_{H})\td y_{+}\td y_{-}\td y_{H} \, \td z \,,
\end{align}
with $-1 \leq z \leq 1$, $0 \leq y_{i} \leq 1$, and $z$ being the cosine of an angle between the initial (anti)particle and the final particle three-momenta.  Two angular degrees of freedom have been integrated out. The asymptotic cross section is consequently
\bea
	\sigma &=&\frac{\Big(2\log\left(s/m_{t}^{2}\right) - 1\Big)\delta_{t}^2
	+	2 \delta_{t} \delta_{Z} + \frac{1}{4} \delta_{Z}^2}{64 (2\pi)^{3}v^2} 
 \, ,
\label{intM2}
\eea
where, for left-handed fermions, 
\bea
\delta_{t} &=& \frac{2 m_t^{2}}{v^{2}}\,\left(C_t-C_W\right) \approx 0.99 \left(C_t-C_W\right),
\nonumber\\
\delta_{Z} &=& \frac{2 M_Z^{2}}{v^{2}}(2q_b s_w^{2} + 1)\left(C_{Z} - C_{W}\right) \approx 0.23\left(C_Z-C_W\right)\, ,
\nonumber
\eea
while for right-handed fermions $\delta_{t} = 0$, and in $\delta_{Z}$ the expression $2q_b s_w^{2} + 1$ is replaced by $2\,q_b s_w^{2}$.
The dominant contribution arises from the $\delta_{t}$ terms.

Given the above cross section, the unitarity bound  can now be obtained by requiring \cite{Dicus:2004rg}
\bea
\sigma \lesssim \frac{4 \pi}{s}
\label{eq:bound}
\eea
that holds under the assumption that the $s$-wave contribution dominates the elastic $b\bar b \to b\bar b$ scattering. The above inequality provides the tightest bound that perturbative unitarity can cast.

In order to simplify the analysis, we consider now two different scenarios for Higgs anomalous couplings. We first assume a universal rescaling of the Higgs couplings to weak gauge bosons  by imposing $C_Z = C_W=C_V$. Secondly, we assume $C_f=C_W$ for the Higgs fermion couplings, and vary the relative strength of $C_Z$ and $C_W$,  inducing in this way an explicitly breaking of the custodial symmetry.

Then we obtain
\begin{itemize}
\item if $C_Z = C_W \equiv C_V$ (\ie $\delta_Z=0$ in Eq.(\ref{intM2})),  the bound in Eq.~(\ref{eq:bound}) is given by
\bea
\frac{4\pi}{s} &\geq&  \frac{m_t^{4}\left(C_t-C_W\right)^{2}}{16(2\pi)^{3}v^{6}} \left(2\log\left(\frac{s}{m_{t}^{2}}\right) - 1 \right)\, ;
\label{bound11}
\eea
\item if $C_t = C_W$  (which sets $\delta_t=0$ in Eq.(\ref{intM2})), breaking the custodial symmetry by setting $C_Z \neq C_W$ yields
\bea
\frac{4\pi}{s} &\geq&  \frac{m_Z^{4}\left(C_Z-C_W\right)^{2}}{64(2\pi)^{3}v^{6}} (2q_b s_w^{2} + 1)^{2} .
\label{bound22}
\eea
\end{itemize}
By defining now the unitarity breaking (UB) energy scale, $E_\UB$, as the specific value of $\sqrt{s}$ for which equalities hold in the above equations, we can see that in the first case $E_\UB$ is minimal when $C_t<0$. In particular, setting $C_{t} = - C_{V} = -1$ we obtain
\bea
E_\UB &\approx& 14 \,\rm{TeV}\, .
\eea
For comparison, a similar value (namely $E_\UB \approx  9.3 \,\mathrm{TeV}$) was found in \cite{Farina:2012xp} for the $Wb \to t H $ partonic process in  
single-top production in association with a  Higgs boson, for $C_W = 1$ and $C_t = -1$. In the second case, given the actual bounds on the ratio $C_Z/C_W$\cite{ATLAS:2013sla}, we can assume at most $|C_{Z} - C_{W}| \sim 0.2$. Correspondingly, the scale of unitarity violation brought by a maximal explicit custodial-symmetry breaking is 
\bea
E_\UB \approx 4700 \,\rm{TeV}\, ,
\eea
that is more than two orders of magnitude higher than the one induced by $C_{t} = -C_{V} = 1$.

In conclusion, we checked that all relevant unitarity bounds are well above the effective $HWW$ partonic c.m. energies  for ${\cal O}(1)$ (or less) variations of the $C_{W,Z,f}$ parameters.
The  partonic cross section for the $HWW$ production at the  LHC collision energies falls indeed in the perturbative regime (and therefore it is safely computable) for the $C_{W,Z,f}$ parameters within currently allowed experimental ranges \cite{ATLAS:2012wma,ATLAS:2013sla,CMS:aya}.

\begin{figure*}
\centering
\begin{minipage}{0.45\textwidth}
\includegraphics[width=\textwidth]{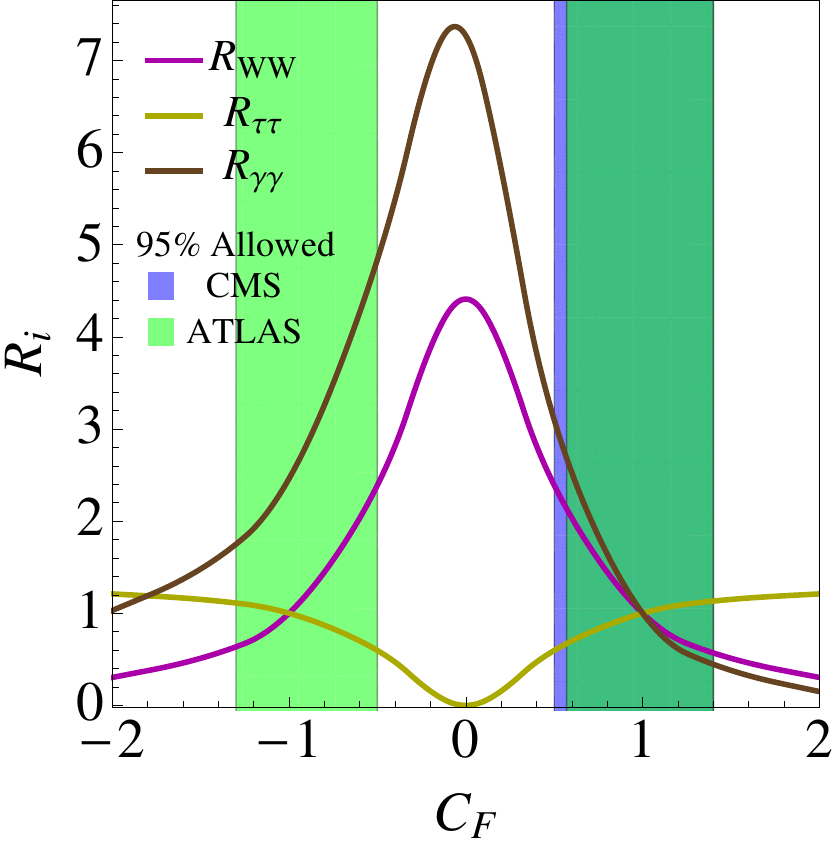}
\end{minipage}
\begin{minipage}{0.45\textwidth}
\includegraphics[width=\textwidth]{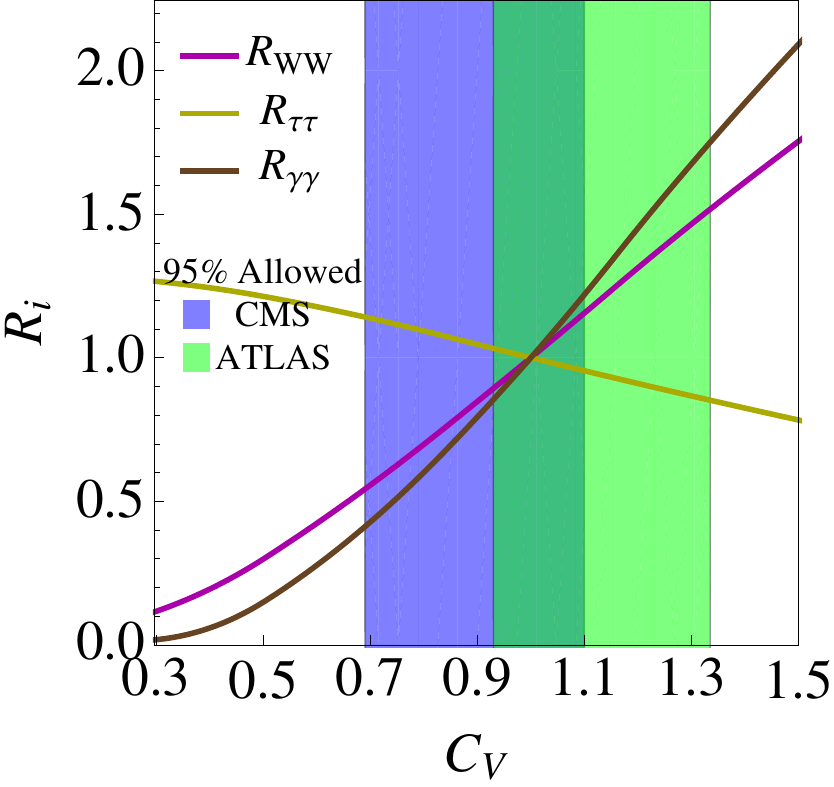}
\end{minipage}
\begin{minipage}{0.45\textwidth}
\includegraphics[width=\textwidth]{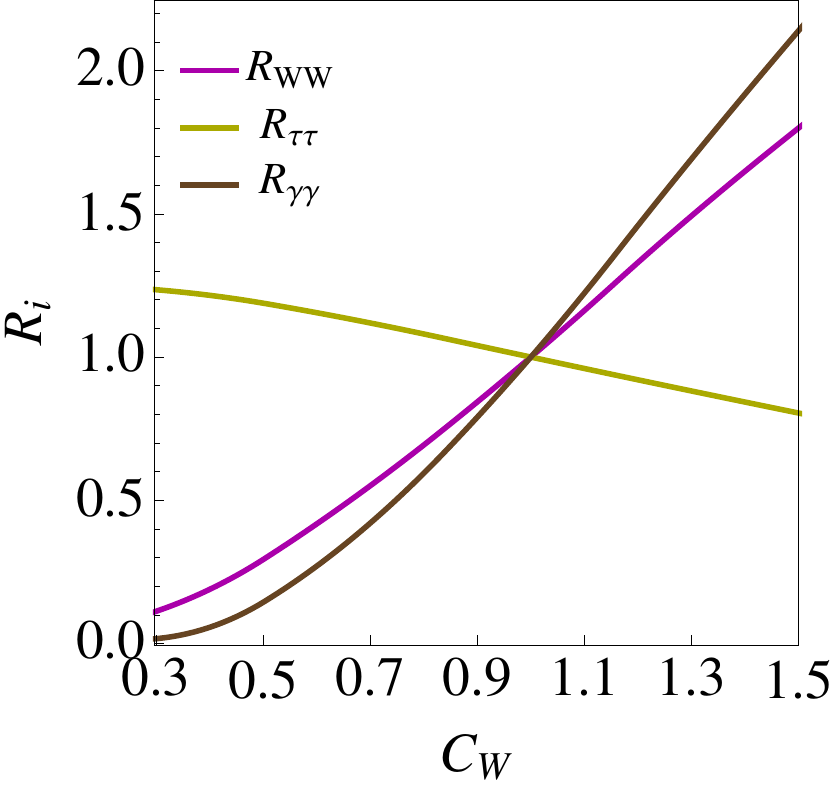}
\end{minipage}
\begin{minipage}{0.45\textwidth}
\includegraphics[width=\textwidth]{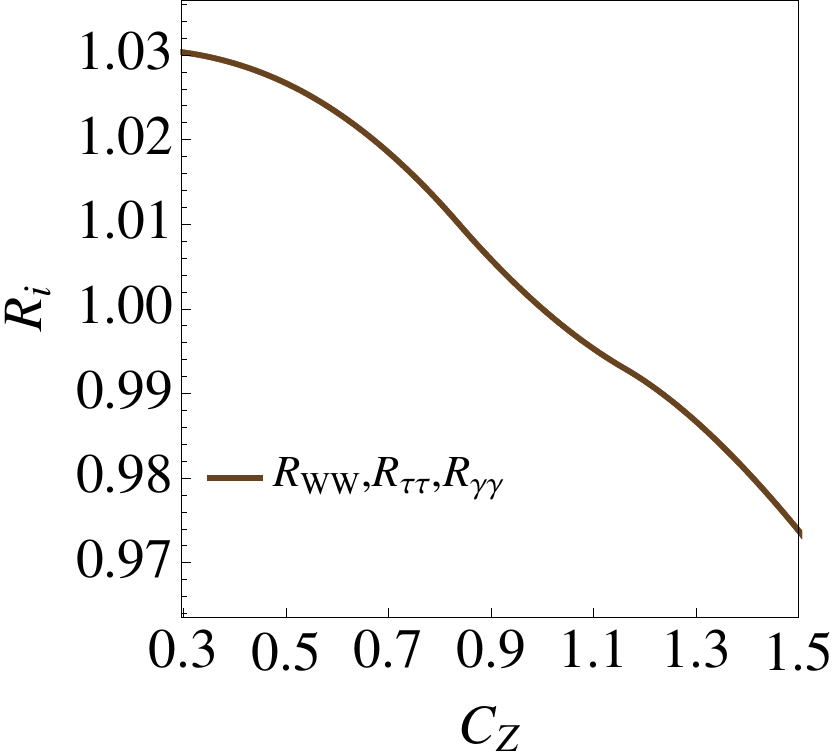}
\end{minipage}
\caption{Higgs boson BR's, normalized to their SM value, as a function of   $C_F$, where $F=t,b,c,\tau ...$, for $C_V=1$ (upper left panel),  $C_V$, where $V=W,Z$, for $C_F=1$ (upper right panel), $C_W$, for  $C_{Z,F}=1$  (lower left panel), and $C_Z$, for  $C_{W,F}=1$  (lower right panel). Here, $R_{i}=BR_{i}/BR_{i}^{\rm SM}$, where  $BR_{i}=\Gamma(H\rightarrow i)/\Gamma(H)_{\rm tot}$. The normalized BR's for $H\rightarrow W^+W^-$, $H\rightarrow \tau\tau$ and $H\rightarrow \gamma\gamma$ are shown by the magenta, yellow and brown lines, respectively.
 In the upper plots, the blue and green areas show the regions allowed at 95\% confidence level by the CMS and ATLAS experiments, respectively.}
\label{fig:BRs}
\end{figure*}
\begin{figure*}[t]
\centering
\begin{minipage}{0.45\textwidth}
\includegraphics[width=\textwidth]{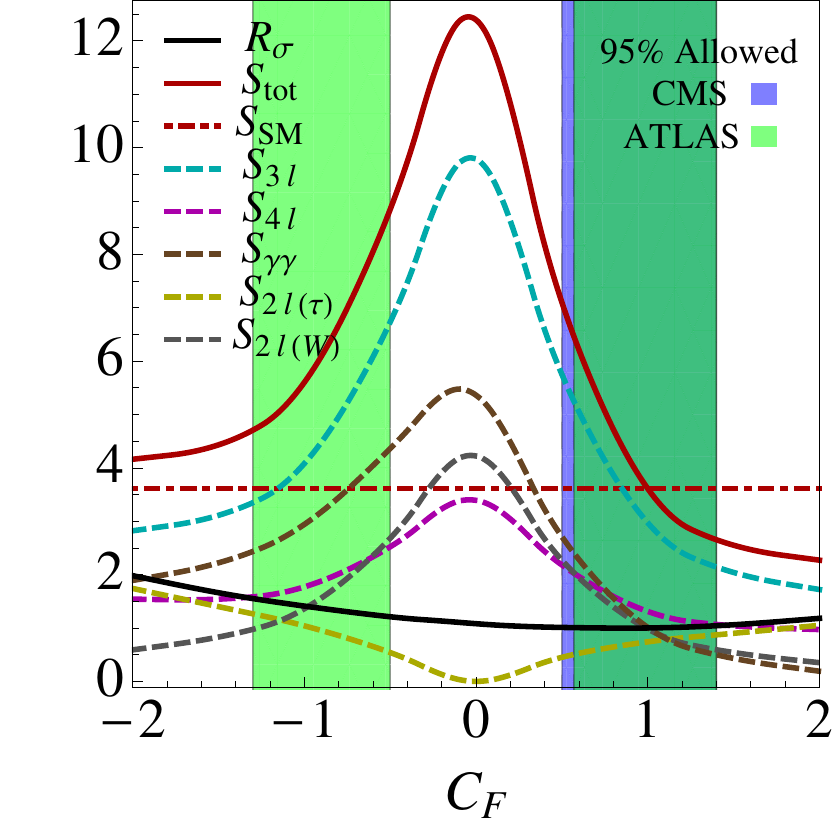}
\end{minipage}
\begin{minipage}{0.45\textwidth}
\includegraphics[width=\textwidth]{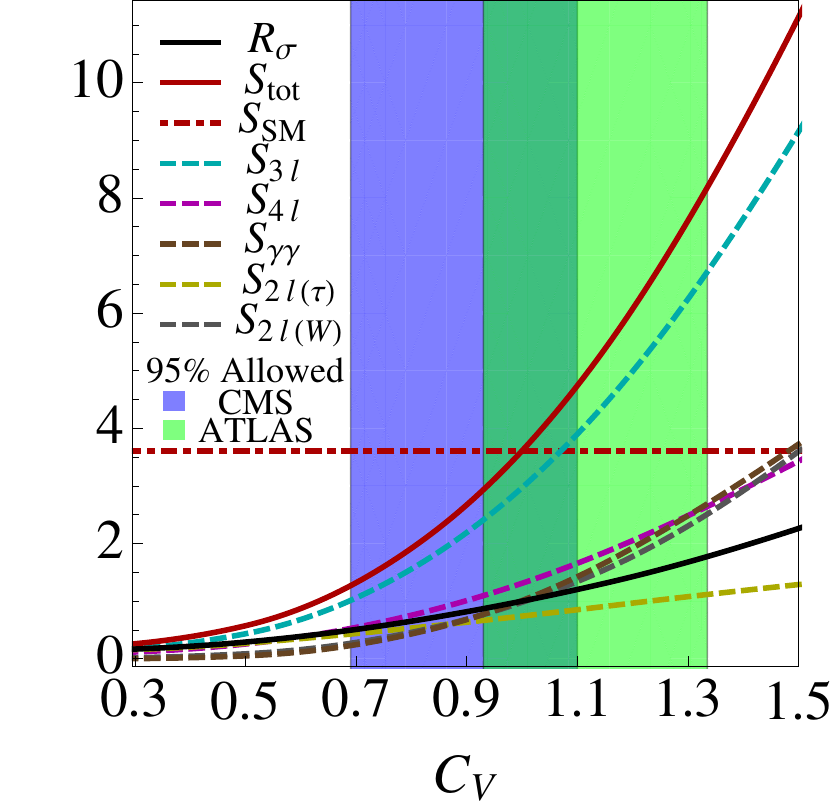}
\end{minipage}
\begin{minipage}{0.45\textwidth}
\includegraphics[width=\textwidth]{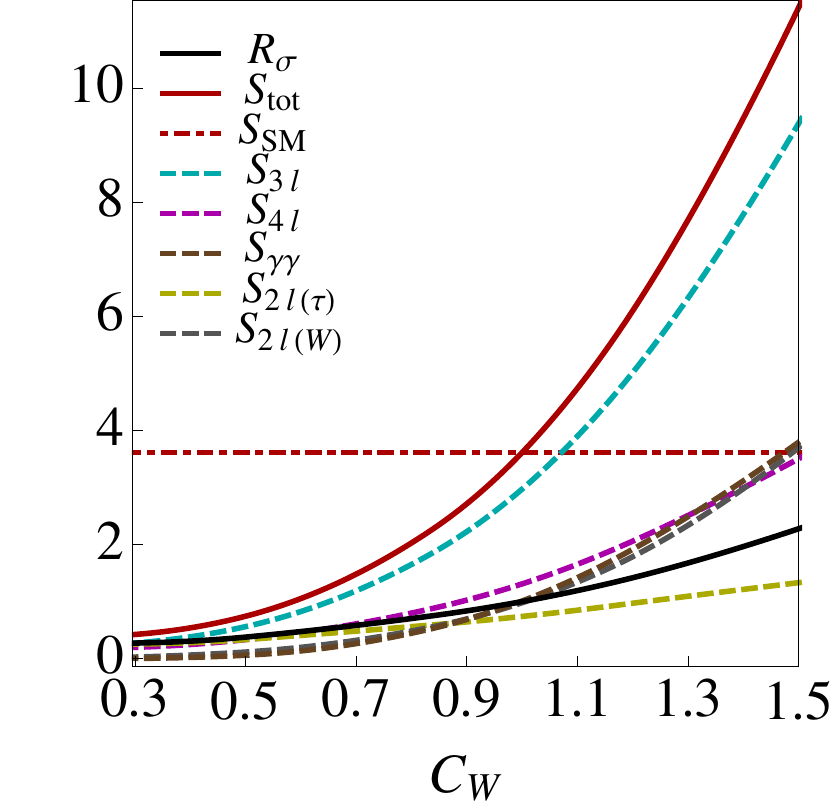}
\end{minipage}
\begin{minipage}{0.45\textwidth}
\includegraphics[width=\textwidth]{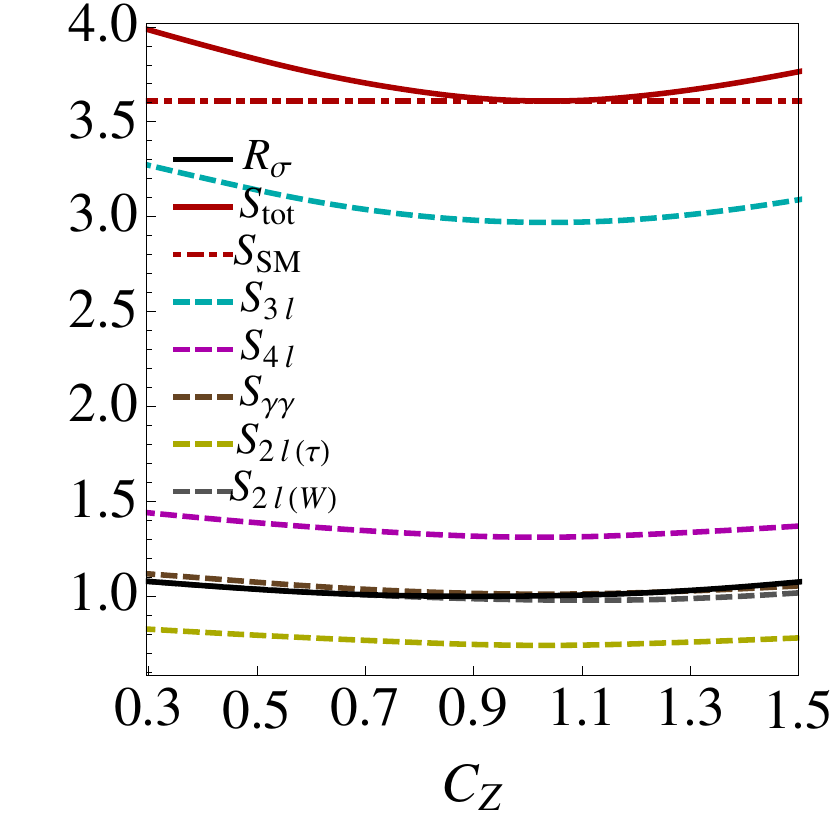}
\end{minipage}
\caption{The  $pp\rightarrow WWH$ cross section, normalized to its SM value, $R_{\sigma}=\sigma/\sigma_{SM}$  at 14 TeV (black solid line), and the combined signal significance (red solid line), corresponding to an integrated luminosity of 3000 ${\rm fb}^\mo$, 
as a function of   $C_F$,  for $C_V=1$ (upper left panel),  $C_V$, where $V=W,Z$, for $C_F=1$ (upper right panel), $C_W$, for  $C_{Z,F}=1$  (lower left panel), and $C_Z$, for  $C_{W,F}=1$  (lower right panel).
 In the upper plots, the blue and green areas show the regions allowed at 95\% confidence level by the CMS and ATLAS experiments, respectively. The individual significances of the five final states [$4\ell,3\ell,\gamma\gamma$,$2\ell(\tau\tau)$,$2\ell(WW^\star)$], in units of standard deviations $\sd$, are shown by the dashed magenta, cyan, brown, yellow and gray lines, respectively. The horizontal red dot-dashed line shows for reference the combined signal significance in the SM.}
\label{fig:xsections}
\end{figure*}
\subsection{Signal strengths and  significances}  
We now discuss the sensitivity of the different $pp\to HWW$ channels analyzed in Section III  to presently allowed variations of the $C_{W,Z,f}$ parameters. 
\\
We first review the impact of such variations on the Higgs BR's. Then, we combine the latter information with
the $pp\to HWW$ cross-section dependence on $C_{W,Z,f}$, obtaining in this way the sensitivity of 
production rates and significances to anomalous Higgs couplings for different $HWW$ signatures.

In the following analysis, we assume that  Higgs couplings  to all fermions are modified by a  universal rescaling coefficient $C_F$, defined as $C_F=C_f$ for all fermions $f$.
As above, $C_V$ is defined as a common rescaling factor for  $g_{WWH}$ and  $g_{ZZH}$, namely $C_{V}=C_{W}=C_{Z}$.

In Figure \ref{fig:BRs} we plot  the Higgs BR's 
normalized to their SM values ($R_i=BR_{i}/BR_{i}^{\rm SM}$), 
for the decay channels relevant to our
analysis, namely $H\to \gamma  \gamma$, $H\to WW^*$, $H\to \tau \tau$,  as a function of anomalous couplings in the range
\bea
-2 < C_F < 2\; , ~~~~~~~ 0.3 < C_{W,Z,V} < 1.5\; .
\eea
In particular, in the top panels of Figure \ref{fig:BRs}, we plot the normalized BR's, $R_i$, 
versus $C_F$,  for $C_V=1$ (left), and $C_V$, for $C_F=1$  (right),  while in 
the bottom panels, the same quantities are plotted versus $C_W$, for $C_Z=C_F=1$   (left), and $C_Z$, for $C_W=C_F=1$  (right). 
The blue and light-green areas, in the top panel plots, label the regions allowed at 95\% C.L. by the present CMS and ATLAS analysis, respectively, where the 
darker-green areas stand for their overlaps\footnote{In the bottom plots of Figure \ref{fig:BRs}, we do not report the experimental allowed regions, since these correspond to a different hypothesis with respect to the one used for the 
exclusion regions of couplings in the $C_F/C_V$ plane adopted by CMS and ATLAS analysis.} \cite{ATLAS:2012wma,ATLAS:2013sla,CMS:aya}.

One can see from the upper-left plot in Figure \ref{fig:BRs} that the BR's for $H\to \gamma\gamma$ and $WW^*$ ($H\to \tau \tau$) reach their maximum (minimum)  at $C_F=0$, which makes the Higgs total width minimal. 
Notice that the maximum of  $R_{\gamma\gamma}$ is not set exactly at $C_F=0$, since the corresponding decay width is not symmetric under a change in the $C_t$ sign. This is due to the destructive (constructive) interference between the $W$- and top-quark  contributions in the $H\to \gamma \gamma$ loop amplitude  for positive (negative) values of $C_F/C_W$.
On the other hand, the positive (negative) slope of $R_i$ 
 for $H\to WW^*, \gamma\gamma$ ($H\to \tau \tau$),
versus $C_{V}$ and $C_{W}$ (in the upper-right and lower-left plots of Figure \ref{fig:BRs}, respectively) is just due to the rescaling property of the $H\to WW^*, \gamma\gamma$ decay widths versus the $C_{V/W}$  coupling. 
In the lower-left plot, we can see that all BR's plotted versus $C_Z$ are degenerate, since the dependence on $C_Z$ mainly affects the total Higgs width (\ie a common normalization factor) in this case.

We now combine the latter results with the $pp\to HWW$  cross-section and signal-rate dependence on Higgs couplings, working  out the potential of the individual five channels analyzed in Section III and their combination.

Figure \ref{fig:xsections} shows, as a function of  anomalous $C_{f,V,W,Z}$, 
  the $pp\to WWH$ total cross section $R_{\sigma}$ normalized to 
its SM value (continuous black line),  the corresponding significance $S_i$, expressed 
in standard deviations ($\sd$), for the five signatures considered in section III A,B,C,D,E (dashed colored lines), and their 
combined effect (continuous red line). 
The horizontal dashed-dot line corresponds to the SM combined significance for the
 five channels. All the significances reported in Figure \ref{fig:xsections} are for a  (14 TeV) LHC integrated luminosity of 3000 ${\rm fb}^\mo$.
 
We checked that in general  the $pp\to HWW$ cross sections grow faster with energy when $C_{f,V}$ depart from the SM set-up, matching the 
expected  unitarity-violation pattern.
The most pronounced effect  is obtained for  negative  top-Yukawa couplings,  $C_t=C_F<0$, that are more sensitive to the unitarity breaking regime
than anomalous $C_Z$, 
as shown by  Eqs.(\ref{bound11}) and (\ref{bound22}).

On the other hand, the cross section dependence on  $C_W$ (lower-left plot in
Figure \ref{fig:xsections}) is mostly a  consequence of the overall 
$C_W^2$ rescaling of the total cross sections, since Higgs radiation from a $W$ boson gives the dominant contribution to the $HWW$ production. Analogous conclusions hold (upper-right plot) for the cross section dependence on a common
$C_V$ rescaling factor.  \\
Quite large variations (up to $50\%$) of the total cross sections are expected for anomalous couplings in the 95\% C.L. range allowed by present experiments. 

We then combine in quadrature the expected $HWW$  significances in different channels,
versus   $C_{V,F}$. Large enhancements can be obtained with respect to the SM signal sensitivity, for $C_{V,F}$ values  
 presently allowed by LHC experiments (see Figure \ref{fig:xsections}), 
thanks to the combined effect of 
the cross section and  BR's dependence on anomalous couplings. In particular, the significance versus $C_F$ for the combined channels is maximal for $C_F\simeq 0$,  
 reaching values up to $\sim 12 \,\sd$, as 
 a consequence of the corresponding  enhancements in the ratios $R_{\gamma \gamma,WW}$  (upper-left plot in Figure \ref{fig:BRs}). For $C_F\sim 0$, the most sensitive  final states are  three-leptons, and  $\gamma \gamma$, followed by 
 two-same-sign leptons. Within the allowed 95\% C.L. regions, the highest  combined significance, corresponding to $C_F\sim -0.5$,  is about $\sim 9 \sd$.

For $C_F=1$, 
 upper-right and lower-left plots in Figure \ref{fig:xsections} give
different-channel significances versus $C_V$ and $C_W$, 
respectively. The  $C_V$ and 
$C_W$ dependence is mainly due to the naive rescaling property of the signal cross section and BR's with $C_V$ and $C_W$. In particular, the maximum  effect,  
corresponding to the largest allowed value $C_V\sim 1.3$, gives a $\sim 8\,\sd$ significance for the combined channels. Analogous conclusions hold for the dependence on $C_W$ (with $C_{Z,F}=1$). 

Finally, in  Figure \ref{fig:xsections}, lower-right plot, we show the   significance versus $C_Z$ (with $C_{W,F}=1$). The maximum enhancement in this case is 
obtained for the lower-edge $C_Z\sim 0.3$, with a significance  $\sim 4\, \sd$, and a modest 10\% enhancement  
over the SM value, 
that falls down to a few per-cent for $0.8 \lsim C_Z \lsim 1.2$.

\section{Summary and Conclusions}
The discovery of the Higgs boson started a new phase in the experimental
 test of the electroweak symmetry breaking mechanism of the SM. Now it is indeed of utmost importance  not only  to study with high accuracy the  Higgs production through the basic discovery  channels, but also to explore lower-cross-section processes
 that can be sensitive to multi-boson interactions. A typical example is given by the Higgs-boson  pair production, which is the lowest-order process that probes at tree-level the trilinear term of the Higgs potential, and yet has a cross section of just 34 fb at the  (14 TeV) LHC. 
 Here we considered the largest-rate  among the electroweak tri-boson production processes involving a Higgs boson in the final state, that is the associated production of a $W$ pair and a Higgs boson. We analyzed (in a tree-level study) the cleanest  experimental signatures corresponding to the $HWW$ final state, that are either multi-lepton or di-photon resonances. The main backgrounds have been scrutinized.
 The most sensitive signature turns out to be a three lepton plus hadronic $W$ final state   
 that reaches a $3\, \sd$ significance at the HL-LHC with 3000 fb$^\mo$.
 Including other channels, we obtain a total $3.6\, \sd$ significance in the SM.
 
 We then carried out a first study of the $pp\to HWW$  sensitivity to possible Higgs anomalous couplings to vector bosons and fermions. We assumed a simple framework where a change in the fermion Higgs coupling sector is universal in fermion flavor.
 Regarding couplings to vector bosons, we assumed both a universal change in the $W/Z$ coupling and the possibility of custodial symmetry breaking.
 
 While the sensitivity to $C_V$ in the cross section is driven by an approximate multiplicative factor $C_V^2$ in the total cross section, the dependence on $C_F$ is  mainly 
 restricted to the $b \bar b \to  HWW$ subprocess, whose amplitude presents, in the SM, non-trivial  cancelation effects between the $W$ and $t$ quark radiation of a Higgs boson.
 
 We also studied unitarity-breaking effects induced by anomalous Higgs couplings in the
 $q \bar q \to  HWW$ amplitude behavior with c.m. scattering energy, and checked that
 the corresponding cross section can be reliably computed at the LHC in the experimentally allowed range of Higgs anomalous couplings.
 
Note that, by the time the high-luminosity run of the LHC will start, our knowledge of Higgs boson couplings will have widely been  enlarged with respect to the present one
\cite{CMS:2013xfa,ATLAS:2013hta}. In case some deviation from the SM expectations in the Yukawa and/or vector boson sectors  will have been observed by then, 
our preliminary study shows that the $HWW$ production mode could be an extra  valuable  channel to clarify the emerging picture. Furthermore, even in a scenario where the SM picture is apparently confirmed, the $HWW$ production could probe higher-dimensional operators by which higher-cross-section processes are moderately affected. We leave to further work the assessment of the $HWW$ potential in this case.
We finally stress that a more reliable evaluation of the $pp \to HWW$ potential for testing the Higgs-boson properties at the LHC will require a more realistic simulation of both theoretical higher-order effects and  the experimental apparatus impact.

\newpage

\mysection{Acknowledgement} We thank M. Raidal for useful discussions. 
E.G. would like to thank the PH-TH division of CERN for its kind hospitality
during the preparation of this work.
This work was supported by the ESF grants 8499, 8943, MTT8, MTT59, MTT60, MJD140, MJD435, MJD298, MJD387, 
by the recurrent financing SF0690030s09 project and by the European Union through the European Regional Development Fund.

\onecolumngrid

\newpage

\appendix

\end{document}